# Two-dimensional MX family of Dirac materials with tunable electronic and topological properties

Yan-Fang Zhang[1,2,†], Jinbo Pan[2,†], Huta Banjade[2], Jie Yu[2], Tay-Rong Chang[3], Hsin Lin[4], Arun Bansil[5], Shixuan Du[1]*, Qimin Yan[2]‡

[1]Institute of Physics & University of Chinese Academy of Sciences, Chinese Academy of Sciences, Beijing, 100190, China

[2]Department of Physics, Temple University, Philadelphia, PA 19122, USA

[3]Department of Physics, National Cheng Kung University, Tainan 701, Taiwan

[4]Institute of Physics, Academia Sinica, Taipei, Taiwan

[5]Physics Department, Northeastern University, Boston, MA 02115, USA

**Abstract**

We propose a novel class of two-dimensional (2D) Dirac materials in the MX family (M=Be, Mg, Zn and Cd, X = Cl, Br and I), which exhibit graphene-like band structures with linearly-dispersing Dirac-cone states over large energy scales (0.8~1.8 eV) and ultra-high Fermi velocities comparable to graphene. The electronic and topological properties are found to be highly tunable and amenable to effective modulation via anion-layer substitution and vertical electric field. The electronic structures of several members of the family are shown to host a Van-Hove singularity (VHS) close to the energy of the Dirac node. The enhanced density-of-states associated with these VHSs could provide a mechanism for inducing topological superconductivity. The presence of sizable band gaps, ultra-high carrier mobilities, and small effective masses makes the MX family promising for electronics and spintronics applications.

The presence of Dirac-cone structures and singularities in the electronic energy spectra of functional materials can endow them with unique properties and promising prospects for both fundamental research and applications.[1] The discovery of graphene, a monolayer honeycomb structure composed of carbon atoms [2], has spurred intense interest in the exotic physics associated with massless fermions, half-integer[3,4]/fractional[5,6]/fractal[7-9] quantum Hall effects (QHE), and other novel phenomena and properties.[10,11] Within the vast space of inorganic 2D compounds,[12-16] graphene,[3,4,17] silicene and germanene,[18] graphynes,[19,20] and other related systems have been predicted to be Dirac materials.[1,21] Dirac cones have been unambiguously confirmed by experiment in graphene. Experimental identification of Dirac-cone structure in silicene and germanene is still controversial,[22-24] where appropriate substrates are needed to preserve the coherence of the Dirac cones. Experimental progress toward identifying Dirac electronic structure in more complex 2D structures, such as the graphynes, is still in its infancy.[25] It is therefore highly desirable to continue the search for novel 2D Dirac systems.

Considering the various requirements associated with crystal symmetries and chemical orbital interactions, it is a challenging task to search for novel 2D systems hosting Dirac cones close to the Fermi energy. Within the framework of tight-binding approximations, the presence of Dirac cone is driven by appropriate combinations of hopping energies,[26,27] which are controlled by details of atomic geometries and the types of 2D crystal lattices involved in a given material. A simple tight-binding analysis shows that a hexagonal cell is the most favorable host for Dirac cones.[28] Notably, both the time-reversal and spatial-inversion symmetries are present in most materials, and provide effective constraints on the Hamiltonian that play a key role in the generation of Dirac-cone structures.[1] Common features in the atomic structures include an even number of atoms in the unit cell and their bipartite nature. These constraints can be used as descriptors for guiding the search for novel 2D Dirac compounds.

In this Letter, utilizing a hypothesis-based discovery process with our recently developed 2D materials electronic structure database (unpublished), we identify a novel class of 2D Dirac compounds with hexagonal lattices. First-principles computations based on density functional theory along with a related tight-binding analysis show that the MX monolayer materials family (M=Be, Mg, Zn and Cd, X = Cl, Br and I) hosts graphene-like band structures with Dirac cones located at the boundaries of the Brillouin zone. Based on atomic orbital analysis, we discuss how

Dirac cones emerge in this class of 2D systems with a charge-unbalanced chemical formula. When spin-orbit coupling (SOC) effects are included, these compounds yield time-reversal-invariant topological insulators. A subset of the MX family is found to host a Van-Hove singularity (VHS) with an enhanced density-of-states close in energy to the Dirac cones, which is often associated with unconventional superconductivity. This offers a potential for the material to host topological superconductivity, such as the *d+id* chiral superconductivity or the odd-parity *f*-wave superconductivity.[29,30] The topologically nontrivial electronic structure revealed here exhibits high tunability and it can be effectively modulated via anion substitution or vertical electric field, suggesting that the MX family would be of interest in developing materials platforms for nano-electronics applications.

We initiated our discovery process from a materials dataset of ~880 compounds in our 2D electronic structure database. Initial structure search was guided by the hypothesis that the Dirac-cone electronic structure is favorable in hexagonal lattices with additional symmetry constraints and crystal-system signatures. We chose bipartite systems with an even number of atoms in the unit cell as favorable factors for producing Dirac cone states in the electronic structure.[1] Only nonmagnetic compounds with inversion symmetry were considered. The candidates that passed through the initial screening were subjected to cation/anion substitutions to enlarge the chemical design space. In this way, our hypothesis-driven search yielded the novel class of MX (M=Be, Mg, Zn and Cd, X = Cl, Br and I) compounds with *p3m1* plane-group to host the Dirac cone structures.

Monolayer MX compounds have an X-M-M-X sandwich structure forming a buckled honeycomb lattice with two sublattices (Fig. 1(a)). Taking the example of monolayer BeCl, we illustrate the interplay between its crystal structure and the appearance of Dirac-cone electronic structure. Each Be atom bonds with three Cl atoms and three Be atoms with Be-Be and Be-Cl bond lengths of 2.61 Å and 2.17 Å, respectively. Based on a Bader charge analysis, charge transfer from a Be atom to the neighboring Cl atoms is 0.93 |*e*|, which is consistent with the large electronegativity difference between Cl and Be (3.16 vs 1.57 on the Pauling scale for Cl and Be, respectively) and indicates an ionic Be-Cl bonding character. Due to an unbalanced chemical formula, the cation-to-anion charge transfer leaves one electron on each Be atom, effectively forming a covalent hexagonal bonding network (Fig. 1(b)).

Band structure of BeCl (Fig. 1(c)) without SOC shows that two bands meet at the Fermi energy at the *K* point. The linear energy dispersion of the two bands in the energy range from -1.0 eV to 0.8 eV relative to the Fermi energy (inset to Fig. 1(c)) indicates that the charge carriers are graphene-like Dirac fermions. The energy dispersion can be described by $E = \hbar v_F k$, where $v_F$ is the Fermi velocity and $k$ is the wave vector. The calculated Fermi velocity in BeCl is $6.64 \times 10^5\ m/s$ for electrons and $4.76 \times 10^5\ m/s$ for holes, which are comparable to the velocities reported in two-dimensional materials with high Fermi velocity.[31-33] No other band is present over the large energy range extending from -3.5 eV to 3.0 eV (relative to the Fermi energy), which is optimal for the detection of Dirac cone as well as for carrier modulation needed in applications. Figure 1(d) shows an overview of the Dirac-cone structures located at the six K points in the Brillouin zone. The band structure of BeCl/*h*-BN (Supplementary Materials Fig. S5) clearly displays a preserved Dirac-cone band on the *h*-BN substrate, indicating the possibility of examining Dirac-cone related exotic phenomena using spectroscopic techniques.

The emergence of Dirac-cone electronic structure is universal in the MX compound family, which we constructed by substituting the cation (Mg, Zn, Cd) or anion atoms (Br, I) with other elements. All these monolayer structures were found to host similar Dirac-cone structures at the Fermi energy as well as ultra-high Fermi velocities, indicating that the Dirac-cone structure originates from the unique occupation of the outermost electronic orbitals in this family. The band structures of MgCl, ZnCl and CdCl are presented in Supplementary Materials Fig. S1(a), and the related ground-state properties of the MX family, including equilibrium lattice constants and the Fermi velocities, are given in Table S1.

With the inclusion of SOC (Fig. 1(e)), the Dirac cone in monolayer BeCl opens up a band gap of 12 meV, which is larger than that in graphene ($0.8 \times 10^{-3}$ meV), but it is comparable to the predicted value in silicene (8.4 meV) and is about half of that in germanene (23.6 meV).[34,35] The Wilson band is an open curve traversing the Brillouin zone (BZ) in the time-reversal invariant plane. Our calculations display an odd number of Wilson bands winding the BZ (Fig. 1(f)), which indicates that the $Z_2$ invariant equals 1. This class of MX 2D compounds thus realize the quantum-spin-Hall insulators state.

In the MX compound space, layered ScCl,[36] ZrBr,[37] and ZrCl[38] with the same atomic structure have been successfully synthesized. Electronic structures of ScCl, ZrBr, ZrCl have been studied computationally.[39] 2D ScCl monolayer has been predicted to host intrinsic ferromagnetism.[40] The phonon dispersion was calculated to estimate its dynamical stability as summarized in the Supplementary Materials Fig. S1(b). The absence of imaginary modes in the entire BZ indicates the dynamic stability of these monolayers. The calculated cohesive energy of BeCl (alpha-BeCl) is 3.15 eV/atom, comparable to another phase of BeCl (beta-BeCl) monolayer (~3.27 eV/atom), which has been predicted as a semiconductor with a tunable band gap and good stability,[41] but lower than bulk $BeCl_2$ (3.74 eV/atom). Since the lattice constant of alpha-BeCl is close to that of beta-BeCl, being 3.14 Å and 3.26 Å, respectively, based on our calculations, (it is 3.27 Å for beta-BeCl from reference [44]), it is not clear that a selective growth of alpha-BeCl can be achieved by using different substrates. However, we note that even though this proposed phase is metastable, some of other proposed compounds could possibly be fabricated via controlled growth techniques using different growth times and/or temperatures.[42,43]

We turn next to discuss how the Dirac-cone electronic structure emerges in our proposed MX compounds from the interplay of the frontier orbitals and lattice structure. The projected densities of states (Figs. 2(a), (b)) clearly show that the dominant orbitals contributing close to the Dirac point are the *s*- and $p_z$- orbitals of Be atoms with an *s-p* hybridized nature. The partial charge density between -1 eV and 1 eV (relative to the Fermi energy) represents the dominant contribution from Be atoms (Fig. 2(c)). Furthermore, the Be charge is confined to the region between the two Be layers, which can be attributed to the contribution from Be $p_z$ orbitals. Dirac cones in this family of MX compounds thus originate through a novel combination of unbalanced chemical formulae, orbital hybridization of group-II cations, and sidewise interactions between the two sublattices in a bipartite cation network. The *s* and *p* states in the cation atom form a set of *sp³* orbitals, through which three bonds are formed between the cation and three anions. Due to the unbalanced chemical formula, only one electron of the group-II cation is transferred to the three anion atoms through the σ bonds formed by cation *sp³* orbitals and anion *p* orbitals. One electron remains in the *sp³*-like orbital that points along the *z* direction toward the other sublattice. It is the sidewise overlapping of these *sp³* orbitals from the two sublattices that creates the bonding π and antibonding π* states sandwiched between the two metal layers (Fig. 1(b)). The bonding π band is fully occupied by the remaining electrons, effectively placing the Fermi energy at the Dirac point. To further clarify this picture, we carried out a Crystal-Hamiltonian-Orbital-Population (COHP) analysis of the two states close to the Dirac point (Supplementary Materials Fig. S4). Metal-$s$/metal-$p_z$ and metal-$p_z$/metal-$p_z$ bonding interactions are observed immediately below the Dirac point, while antibonding interactions are presented above the Dirac point. These observations strongly support the novel mechanism we propose for the formation of Dirac cones in the MX family of 2D materials.

Notably, novel topological materials within the 2D MX family have been addressed in a few recent reports.[44-48] We emphasize that the physics underlying the formation of Dirac cones in the family of group-II MX compounds identified in our study differs fundamentally from the work of Zhou *et al.*[44], which focuses on transition-metal compounds with band edges that are constructed via localized d electrons. Dirac cone in Cd-based MX compounds, Yi *et al.* [45] attributed the Dirac cones they found to the interaction between the metal *s*- and anion-*p* states, which does not reflect the actual orbital interaction in these compounds.

Further insight into the nature of the frontier orbitals can be obtained from the computed work functions (Fig. 2(d)), which are strongly correlated with the *s* orbitals of cation species (see detailed explanation in the Supplementary Materials). By hosting a broad distribution of work functions, the MX family offers interesting opportunities for the design of heterojunction-based devices such as the tunneling transistors by taking advantage of the Dirac cone states. With the information of the frontier orbitals close to the Fermi energy in hand, the emergence of the Dirac cone in monolayer BeCl can be effectively understood within our effective tight-binding model, [49,50] which is described in detail in the Supplementary Materials.

The unique band structures of MX compounds also offer a platform for exploring the interplay between topological physics and superconductivity. For example, the band structure of MgCl (Fig. 3(a)) hosts a VHS, which lies close in energy to the Dirac cone. Notably, other MX family members also exhibit the VHS in their electronic band structures (Supplementary Materials Fig. S1(a)), although the VHSs in these other compounds lie farther away from the Fermi energy compared to the case of MgCl. The energy separation in MgCl between the VHS and the Dirac cone is 0.28 eV, which could be modulated by doping and electric field. Figure 3(b) shows the density of states (DOS) with a logarithmic singularity. This indicates the presence of a saddle-point where the massless Dirac cone meets with the pocket centered at M, and presents a transition from an electron-like to a hole-like Fermi surface (Fig. 3(c)) when the chemical potential is raised. Position of a VHS in the reciprocal lattice space is presented in Figure 3d as an example. It has been suggested that the enhanced DOS associated with VHSs could drive unconventional superconductivity. In the electron doped MgCl, there are 6 VHS points located between the M and K points, offering the possibility of odd-parity f-wave order. When considering the hole-doped case, there are 3 VHS points around -1 eV located at the M points. This is similar to the case of graphene and would favor chiral d+id superconductivity. [29,30] The two aforementioned superconducting orders are desirable type of topological superconductivities.

Since the discovery of Dirac-cone structure in graphene, the introduction of a band gap while preserving high carrier mobility has proven to be a challenging task because even a moderate gap opening in graphene is accompanied with a large deterioration of carrier mobility. In contrast, in the monolayer MX compounds, the buckled honeycomb (metal) lattice is intercalated between the two layers of anion atoms, offering unique opportunities for modulating band structures. By inducing an electric potential difference between the atoms in different sublattices and breaking their equivalence, vertical electric field (Fig. 4(a)) is an effective approach for opening a band gap in the vicinity of the Fermi energy.[51,52] Dependence of the calculated band gap in BeCl (with SOC) on the vertical electric field can be seen in Fig. 4(b). Similar to the case of graphene,[53] the electric field gradually closes the band gap until it reaches zero at a critical point. A larger vertical field opens a trivial band gap. By applying a 0.5 V/Å field, monolayer BeCl opens a band gap of ~41.9 meV, which is comparable to that predicted for germanene (~60 meV[51]). [Considering the underestimation of band gaps by DFT, the actual band gap is expected to be larger.]

As another example of band structure modulation, we consider a $Be_2ClI$ structure obtained by substituting one layer of Cl atoms with I atoms (Fig. 4(c)), which in principle could be achieved using doping techniques.[54,55] The optimized lattice constant is 3.47 Å, which as expected is larger than that of BeCl due to the larger atomic radius of I atoms. Band structure of the $Be_2ClI$ (Fig. 4(d)) shows a sizable band gap (0.17 eV using the GGA-PBE functional and 0.22 eV using the HSE hybrid functional) at the *K* point, while the large dispersion of the conduction and valence bands is preserved. Note that this band gap is much larger than the predicted band gap in modulated graphene on *h*-BN, which is only tens of meV.[56] Moreover, we find electronic and valley polarization[57,58] in Be$_2$ClI (Fig. S6), which would suggest its potential use in valleytronics. We estimated the carrier mobilities in Be$_2$ClI by applying a simple phonon-limited scattering model [59]; the related computational details and fitting curves are presented in the *Methods* section and the Supplementary Materials (Table S2, Figs. S2 and S3). The computed carrier mobilities of Be$_2$ClI are on the order of $10^5\ cm^2/V \cdot s$, which are two orders of magnitude larger than those of single- and few-layer black phosphorus.[60] The sizable band gap combined with high carrier mobilities would make Be$_2$ClI a promising candidate material for high-performance radio frequency devices.[61]

In summary, through a hypothesis-based data-driven approach, we predict a promising family of monolayer MX materials, which hosts Dirac-cone electronic structures at high-symmetry k points in the Brillouin zone. VHSs and the corresponding enhanced density of states close in energy to the Dirac point are found for a subclass of the compounds in this family, which offer the potential to host topological superconductivity. We discuss how the Dirac-cone structure originates through an interplay of the bipartite nature of the hexagonal cation network, incomplete cation-to-anion charge transfer, and the sidewise interactions of the hybridized metal orbitals. Our predicted 2D Dirac materials host high Fermi velocities, which are comparable to that of graphene. The presence of a broad distribution of work functions in the MX 2D materials family would provide opportunities for the realization of novel heterojunction-based devices by taking advantage of the Dirac carriers. Electronic structures of MX monolayers are highly tunable via chemical substitutions and vertical electric field, suggesting the viability of these films for developing platforms for various applications.

**Acknowledgement:**

This work was supported by the U.S. Department of Energy, Office of Science, Basic Energy Sciences, under Award #DE-SC0019275. It benefited from the supercomputing resources of the National Energy Research Scientific Computing Center (NERSC), a U.S. Department of Energy Office of Science User Facility operated under Contract No. DE-AC02-05CH11231, and Temple University's HPC resources supported in part by the National Science Foundation through major research instrumentation grant number 1625061 and by the US Army Research Laboratory under contract number W911NF-16-2-0189. Shixuan Du and Yan-Fang Zhang acknowledge support from the National Key Research and Development Projects of China (2016YFA0202300), Strategic Priority Research Program (XDB30000000), National Natural Science Foundation of China (61888102), and the International Partnership Program (112111KYSB20160061) of the Chinese Academy of Sciences.

† Y.F. Z. and J.B. P. contributed equally to this work.

Corresponding authors:

\* sxdu@iphy.ac.cn

‡ qiminyan@temple.edu

**Figure Captions:**

Figure 1. (a) Top and side views of the crystal structure of the MX (M: Be, Mg, Zn and Cd. X: Cl, Br and I) family of compounds. (b) A schematic of Dirac cone formation in MX compounds through unbalanced electron transfer and sidewise hybridized orbitals. (c) Band structure of BeCl without spin-orbit interaction. (d) A 3D view of the valence and conduction bands in the vicinity of the Dirac point. (e) Band structure of BeCl with spin-orbit interaction. (f) Wannier charge-center evolution in BeCl.

Figure 2. (a), (b) Projected density-of-states of BeCl on Be and Cl atoms. (c) Partial charge density of BeCl from -1 eV to 1 eV relative to the Fermi energy, with an isosurface value of 0.034 e/Å$^3$. (d) Work functions, $s$-orbital centers, and $s$-orbital contributions in MX compounds.

Figure 3. (a), (b) Band structure and density-of-states of MgCl. (c) A 3D view and a contour plot of MgCl in the vicinity of the saddle point at $0.38\vec{b_1}+0.38\vec{b_2}$, where $\vec{b_1}$ and $\vec{b_2}$ are the reciprocal lattice vectors. (d) One of the VHS positions in the reciprocal lattice.

Figure 4. (a) Schematic showing the application of a vertical electric field. (b) Band gap of BeCl as a function of the vertical electric field; band gaps are 12.0 meV, 0 meV and 41.9 meV at 0.0 V/Å, 0.118 V/Å and 0.5 V/Å, respectively. (c) Top and side views of Be$_2$ClI. The green, purple and orange balls represent Cl, Be and I atoms, respectively. (d) Band structure of Be$_2$ClI with a gap opening at the Fermi energy.

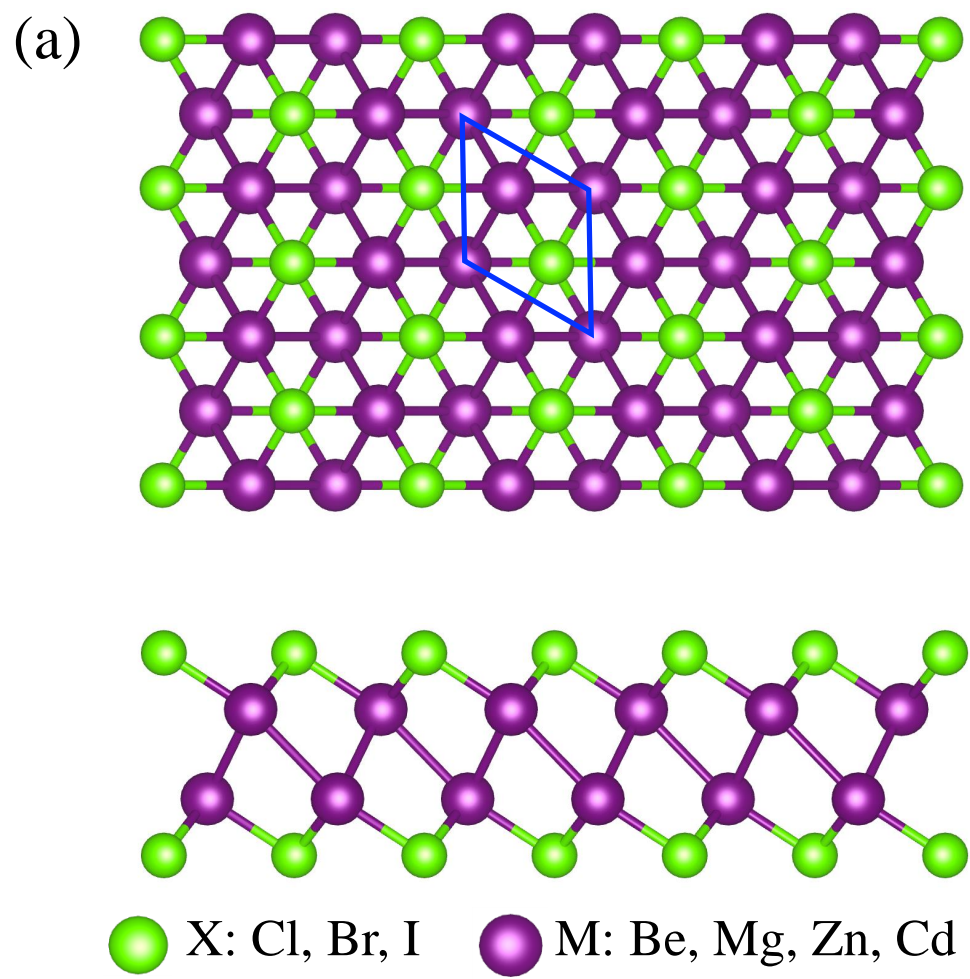
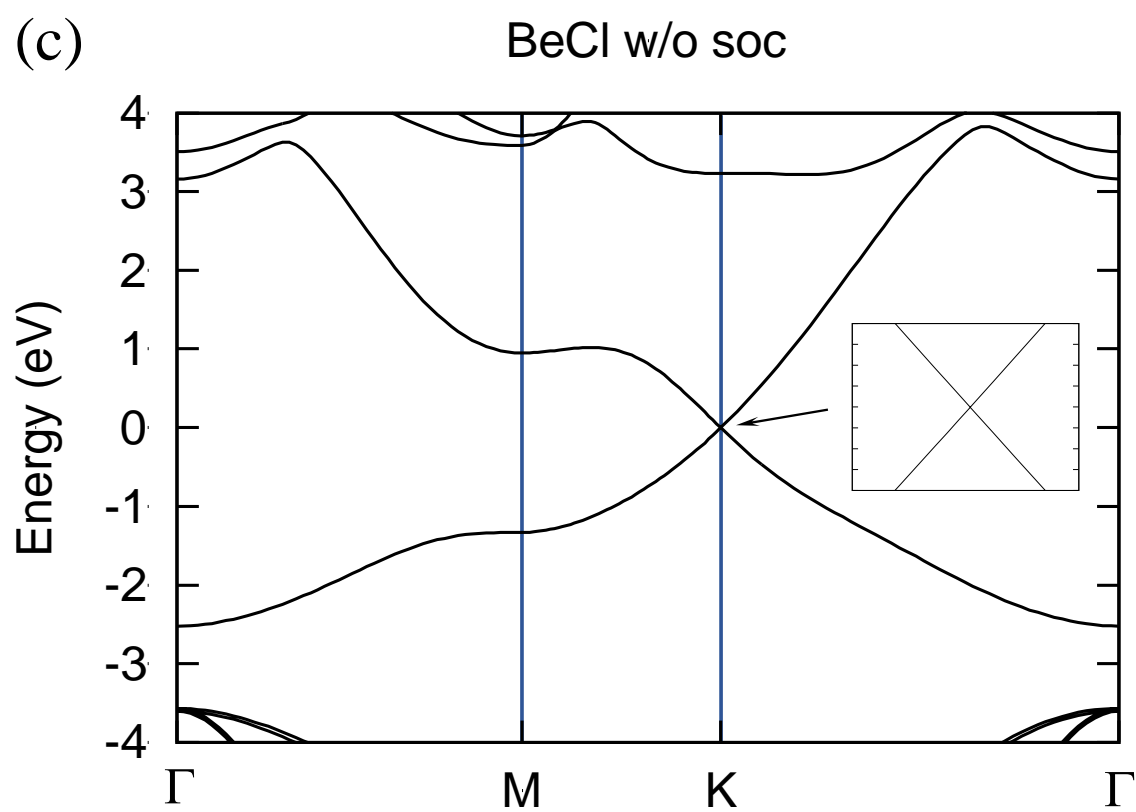
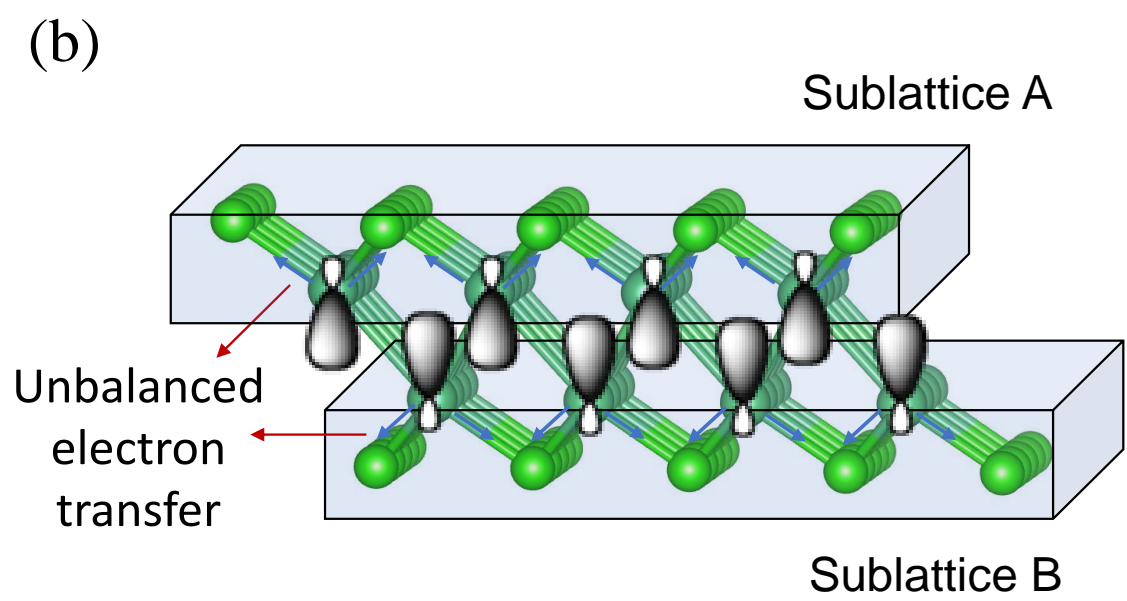
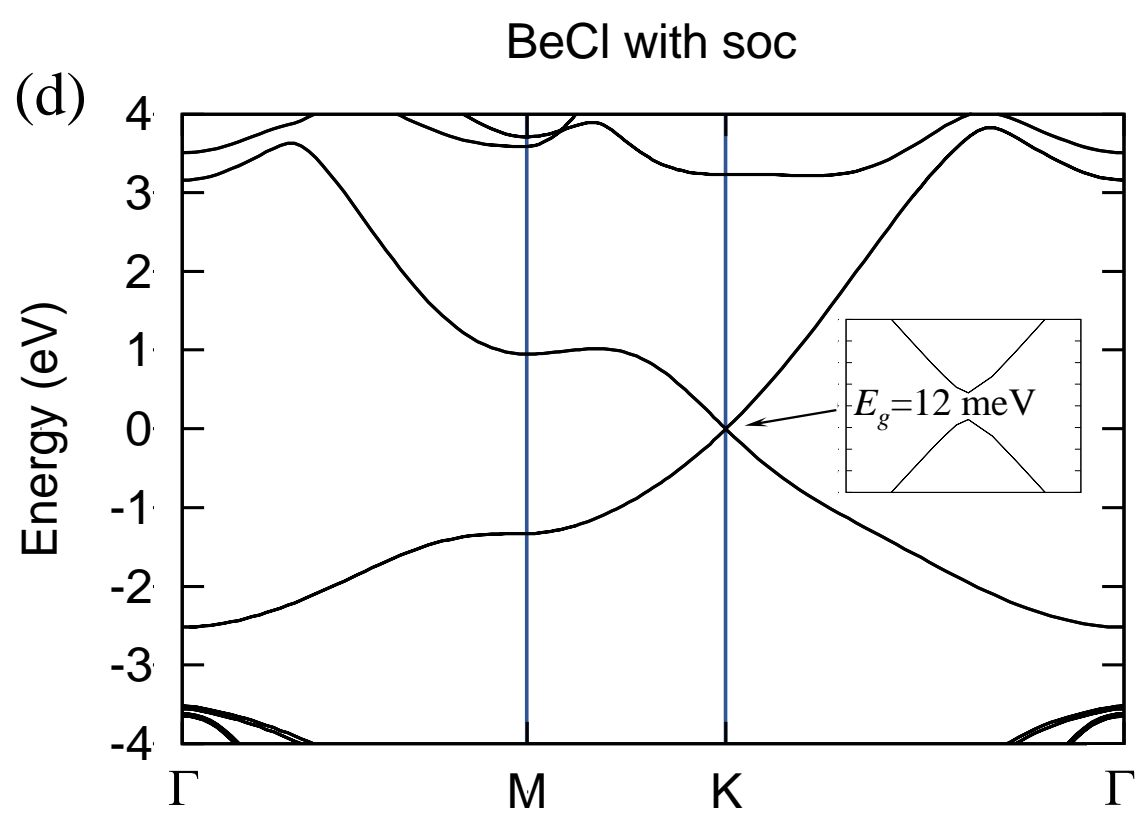
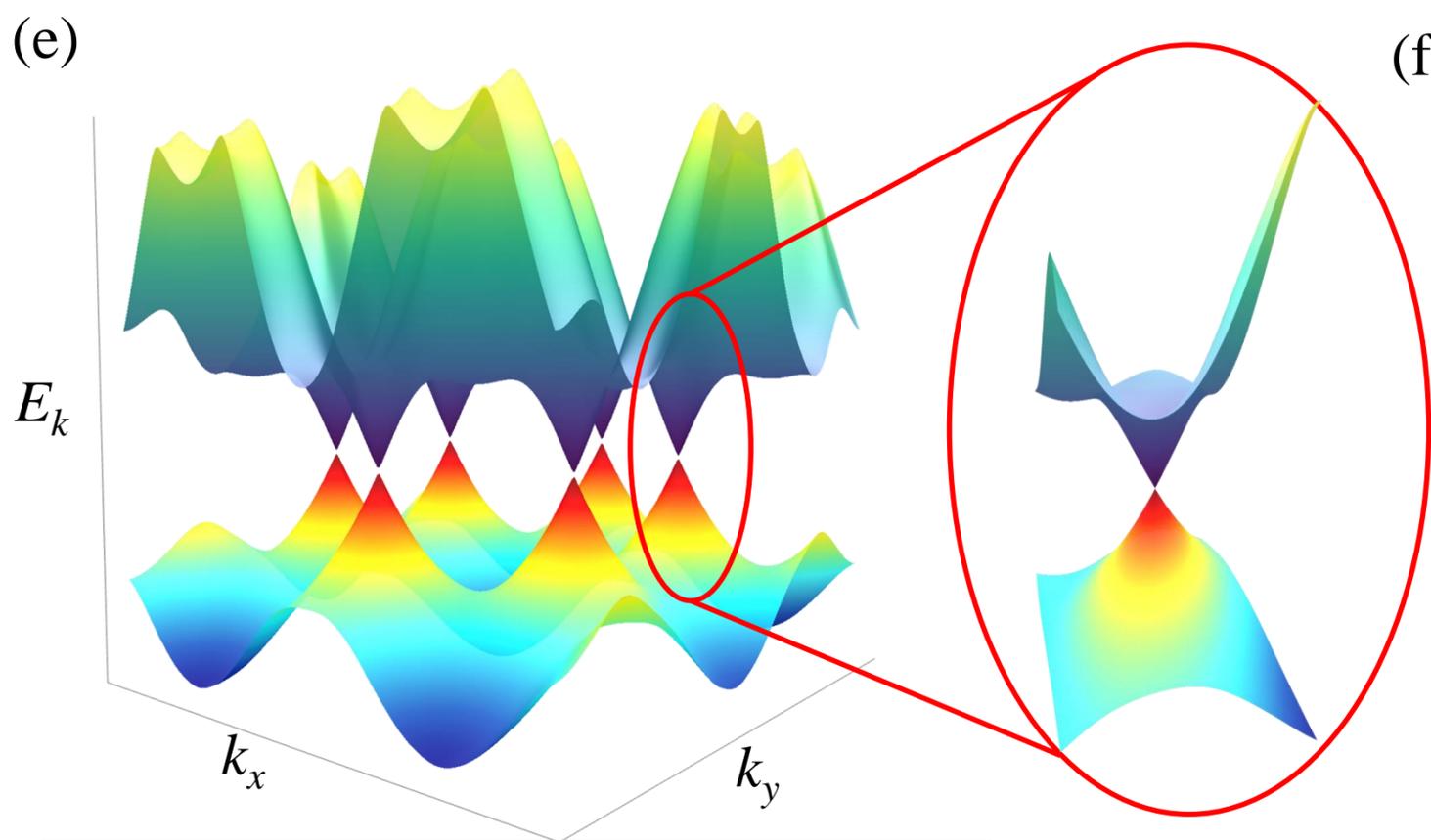
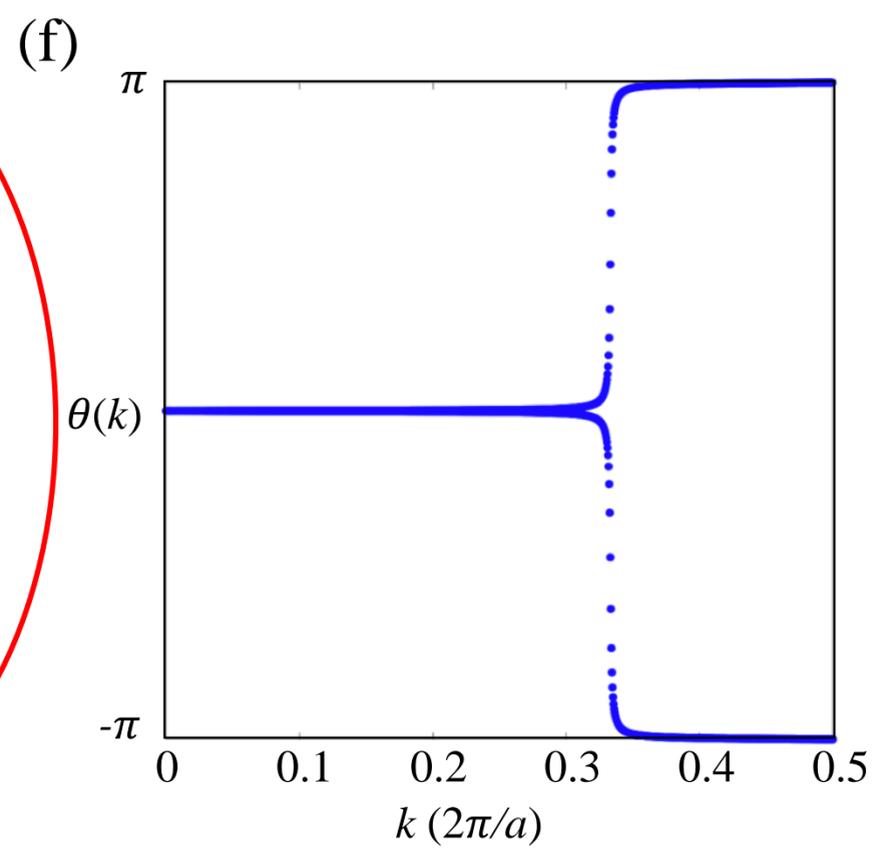

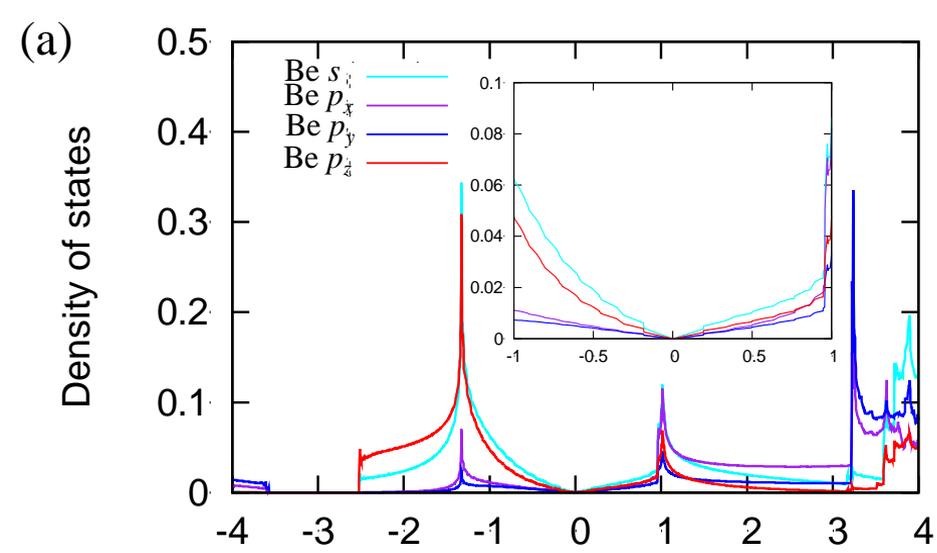
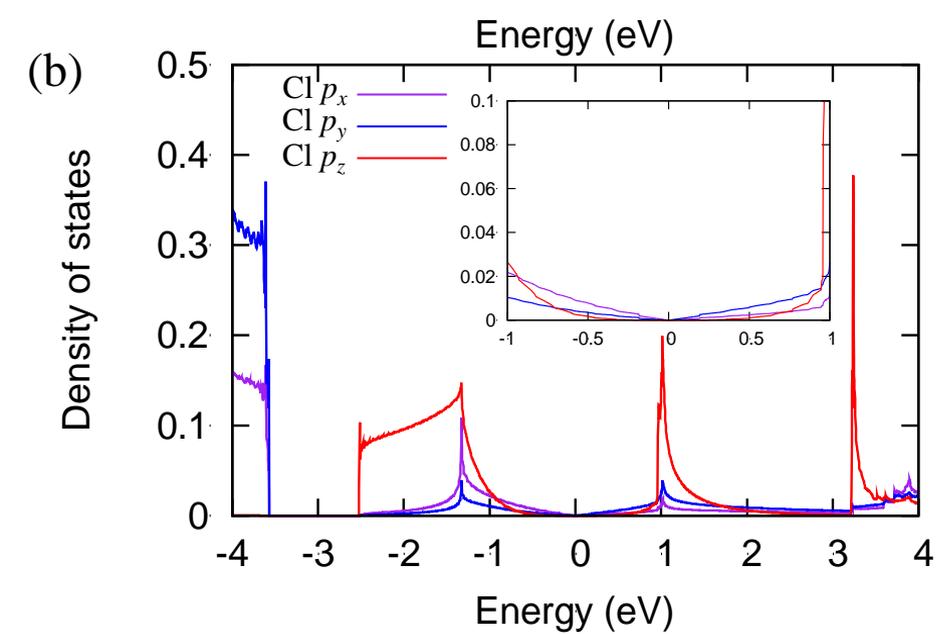
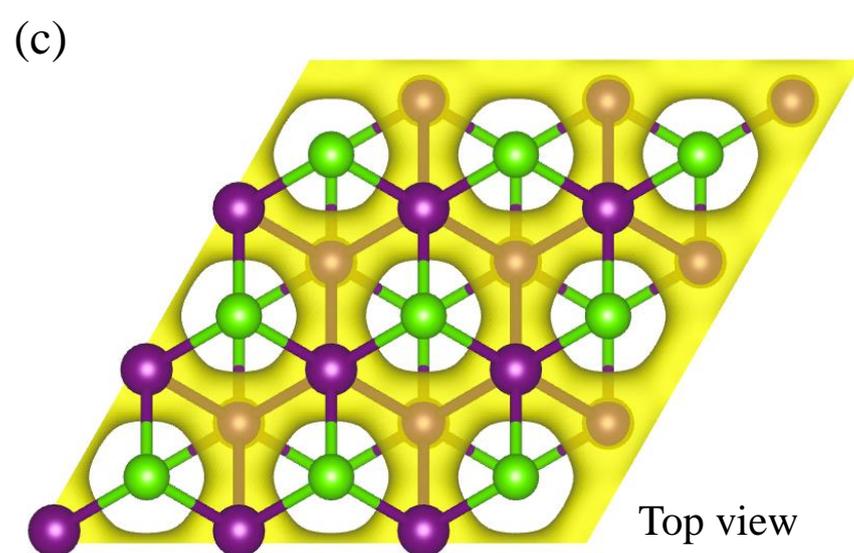
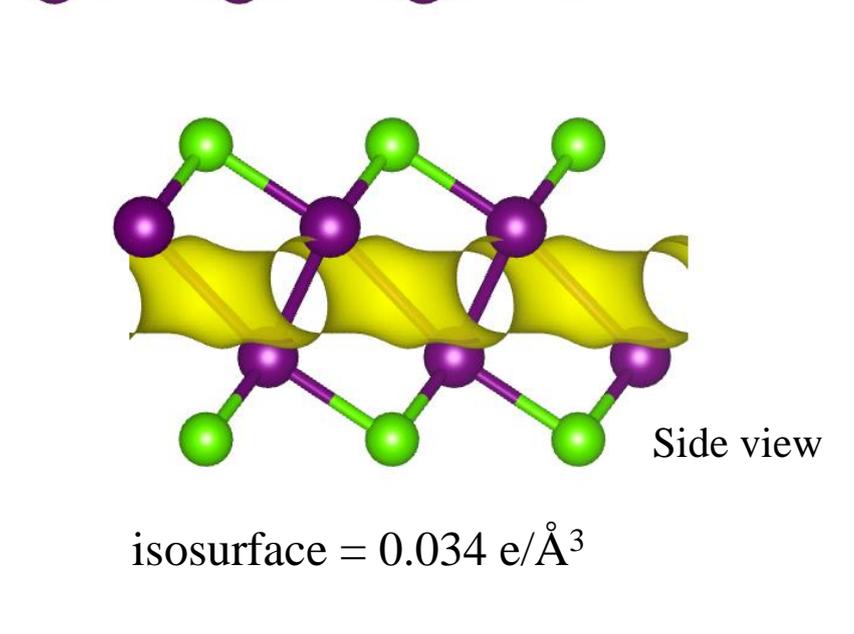
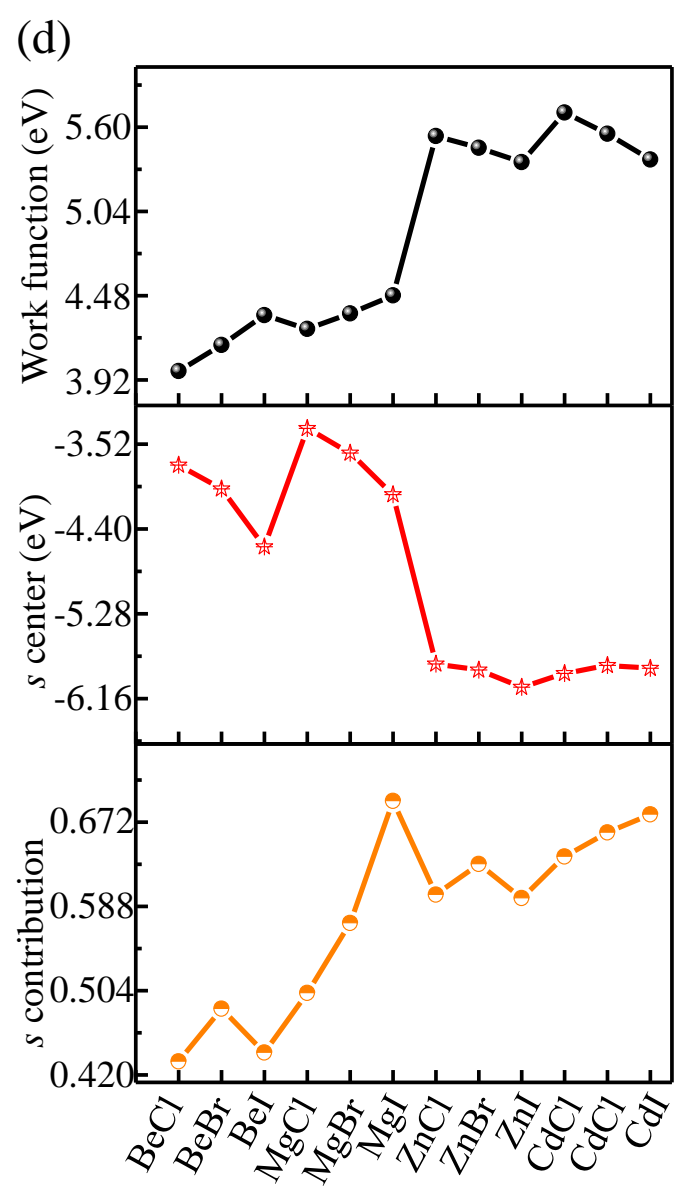

isosurface = 0.034 e/Å³

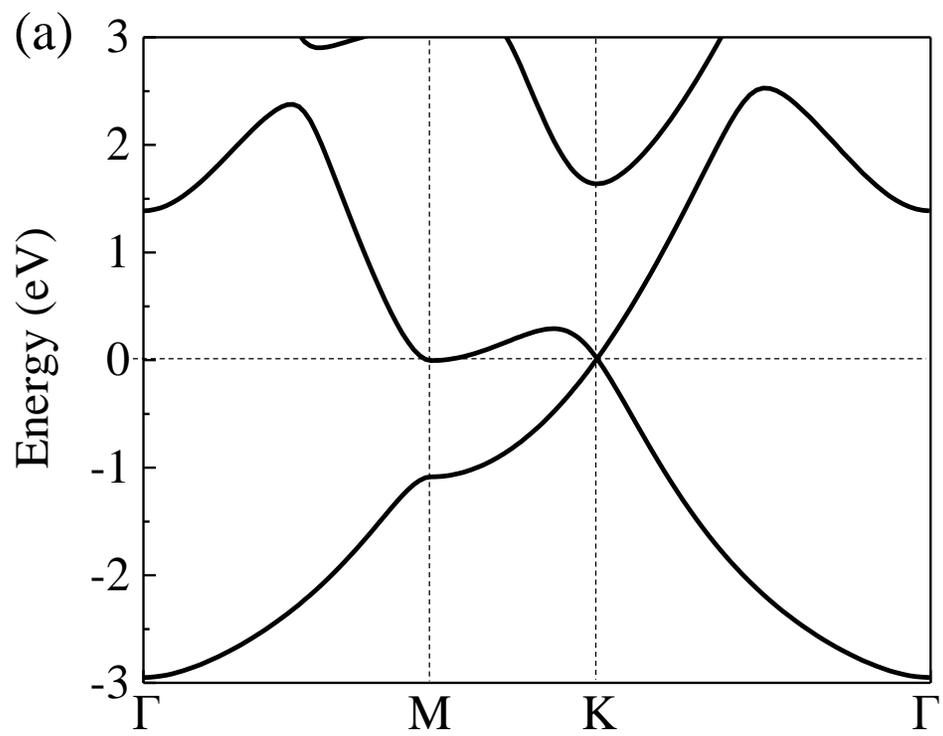
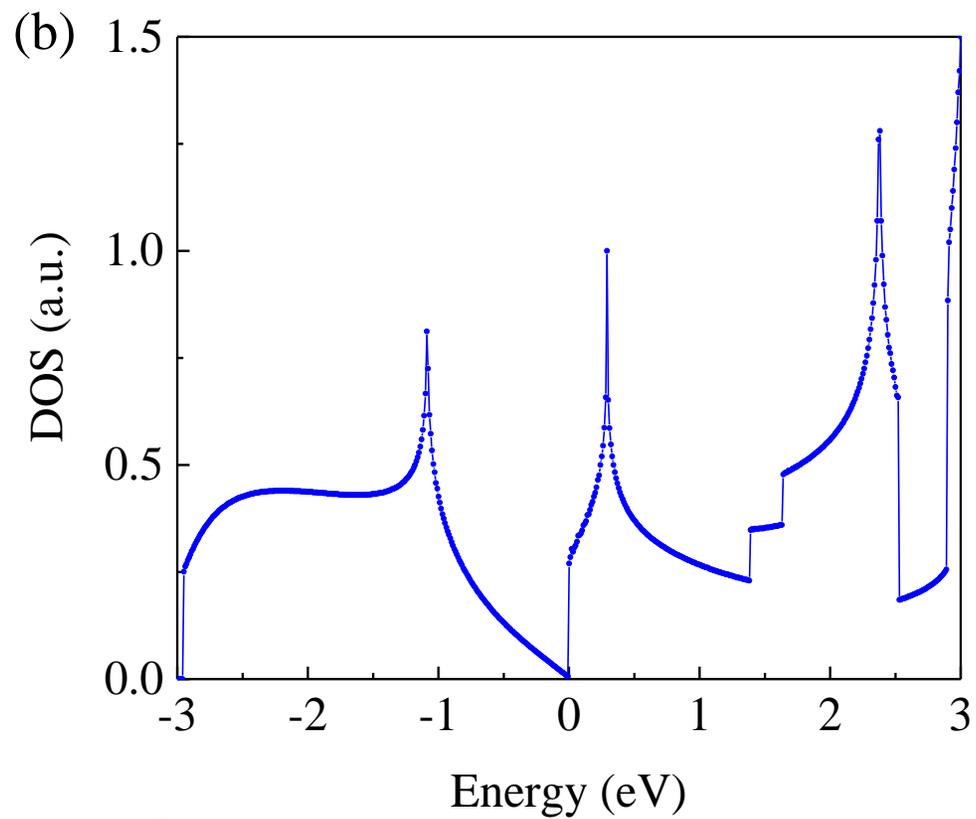
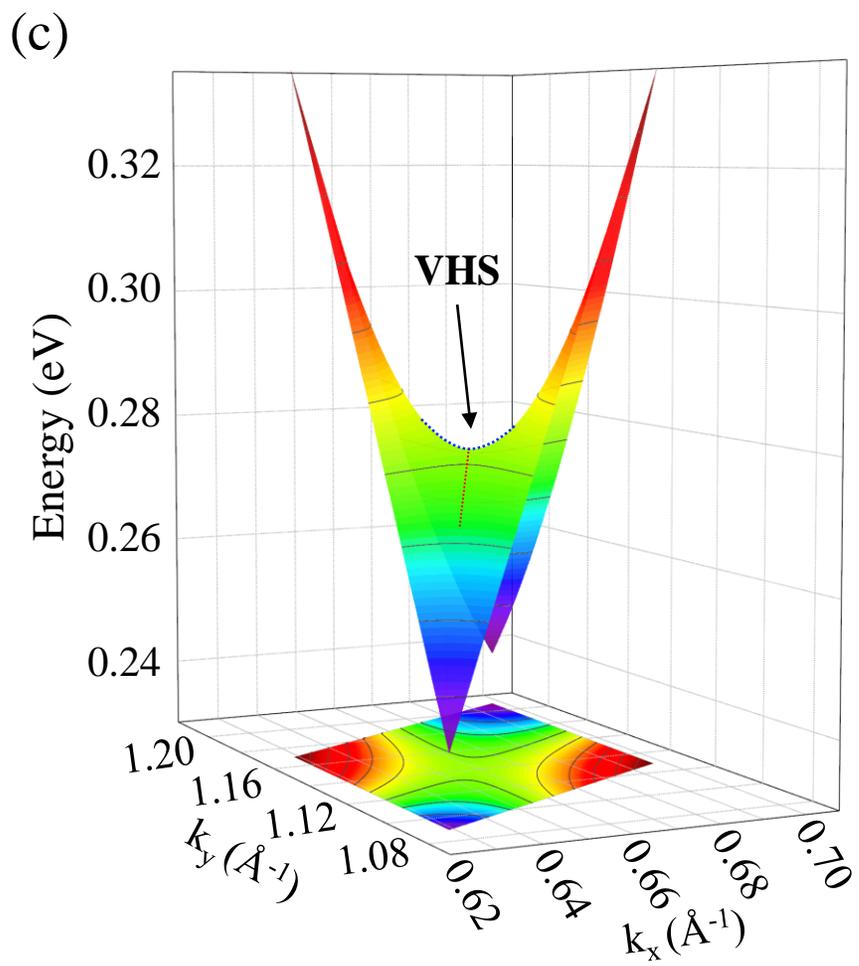
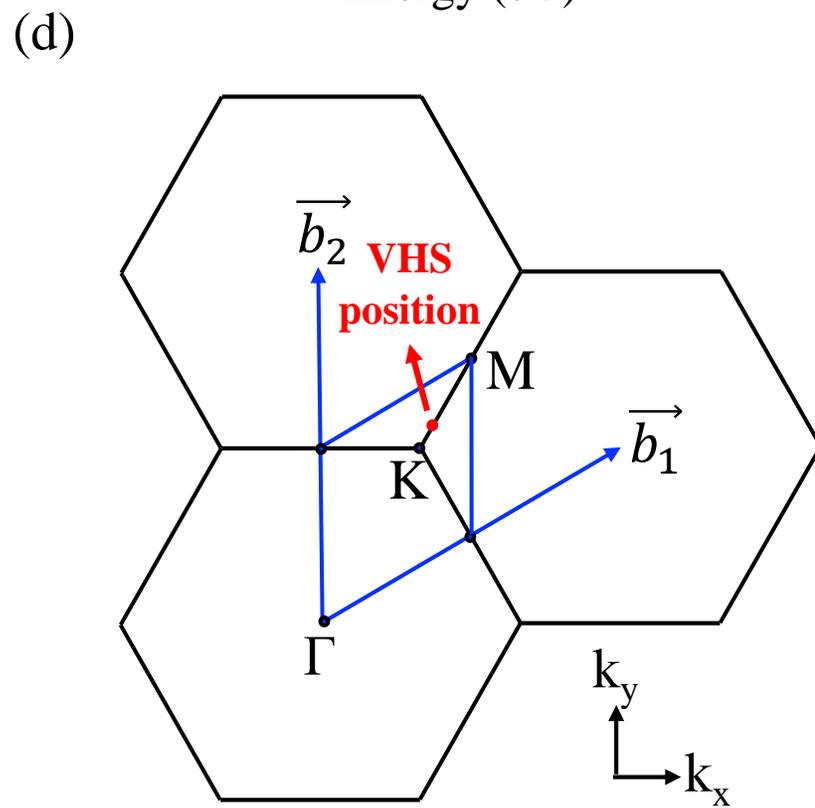

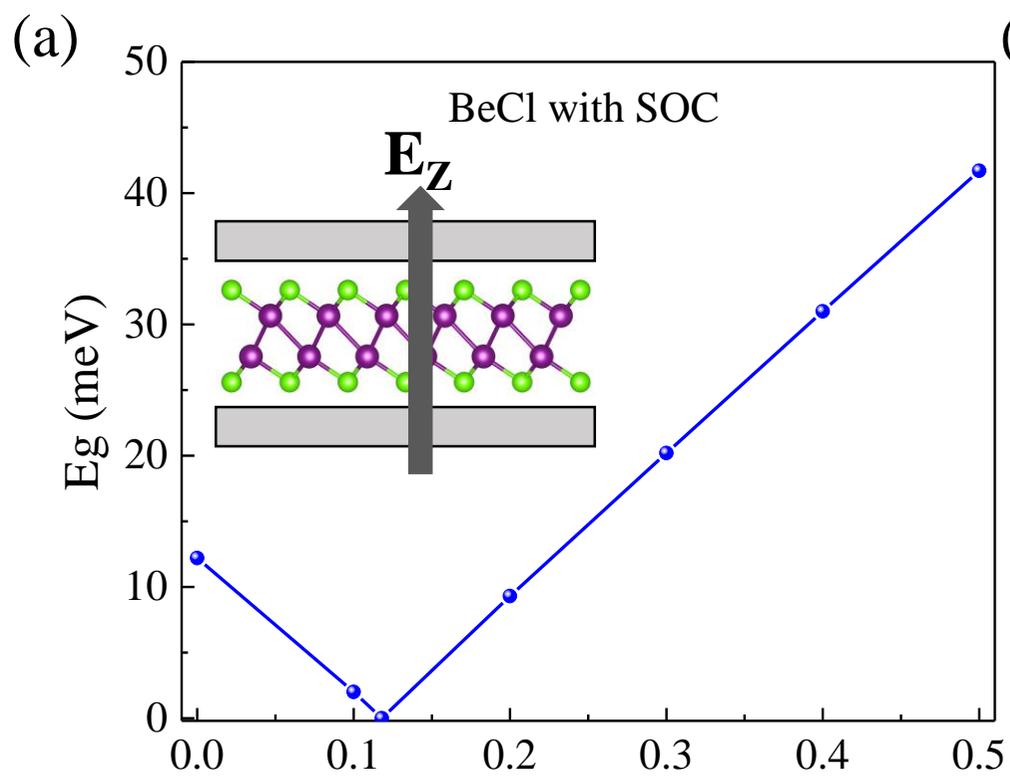
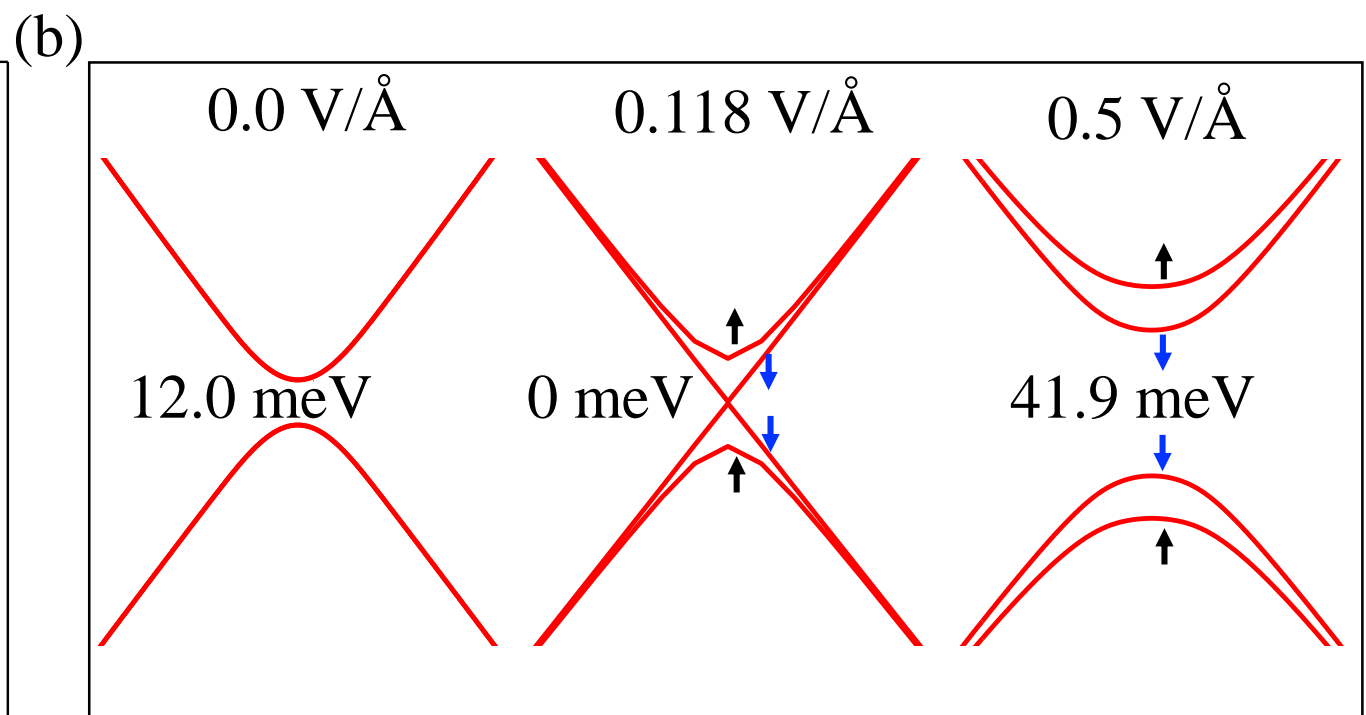
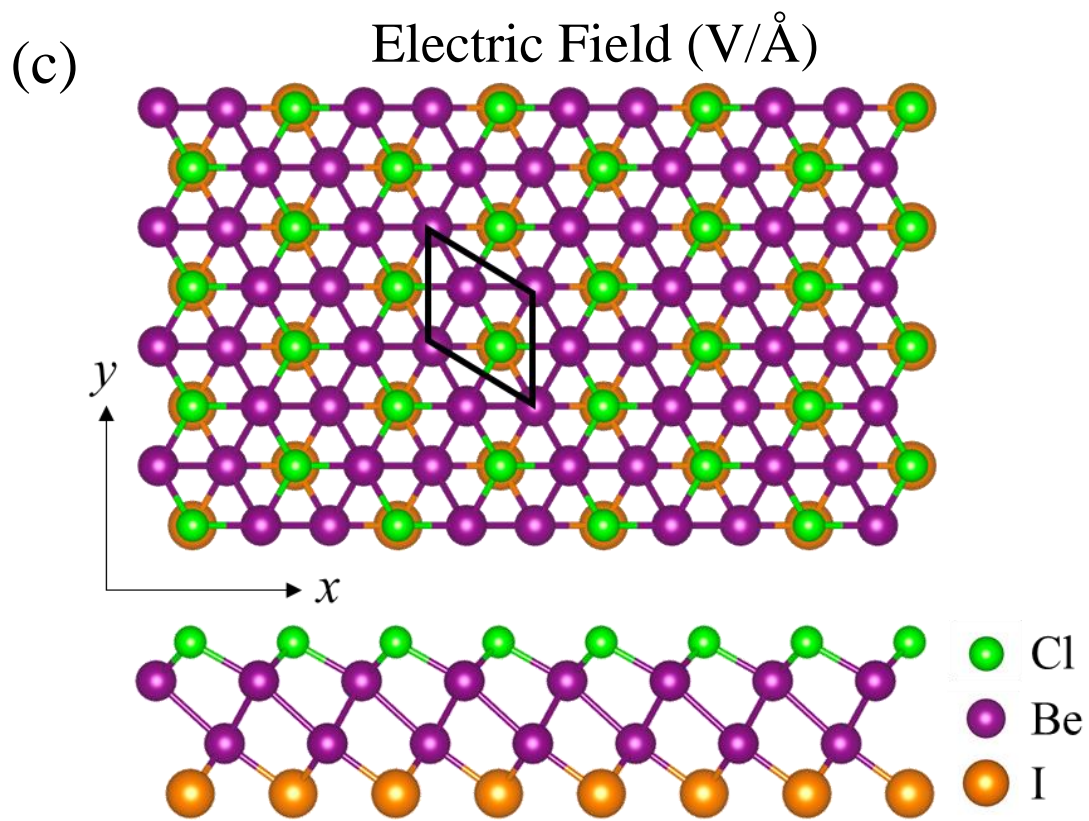
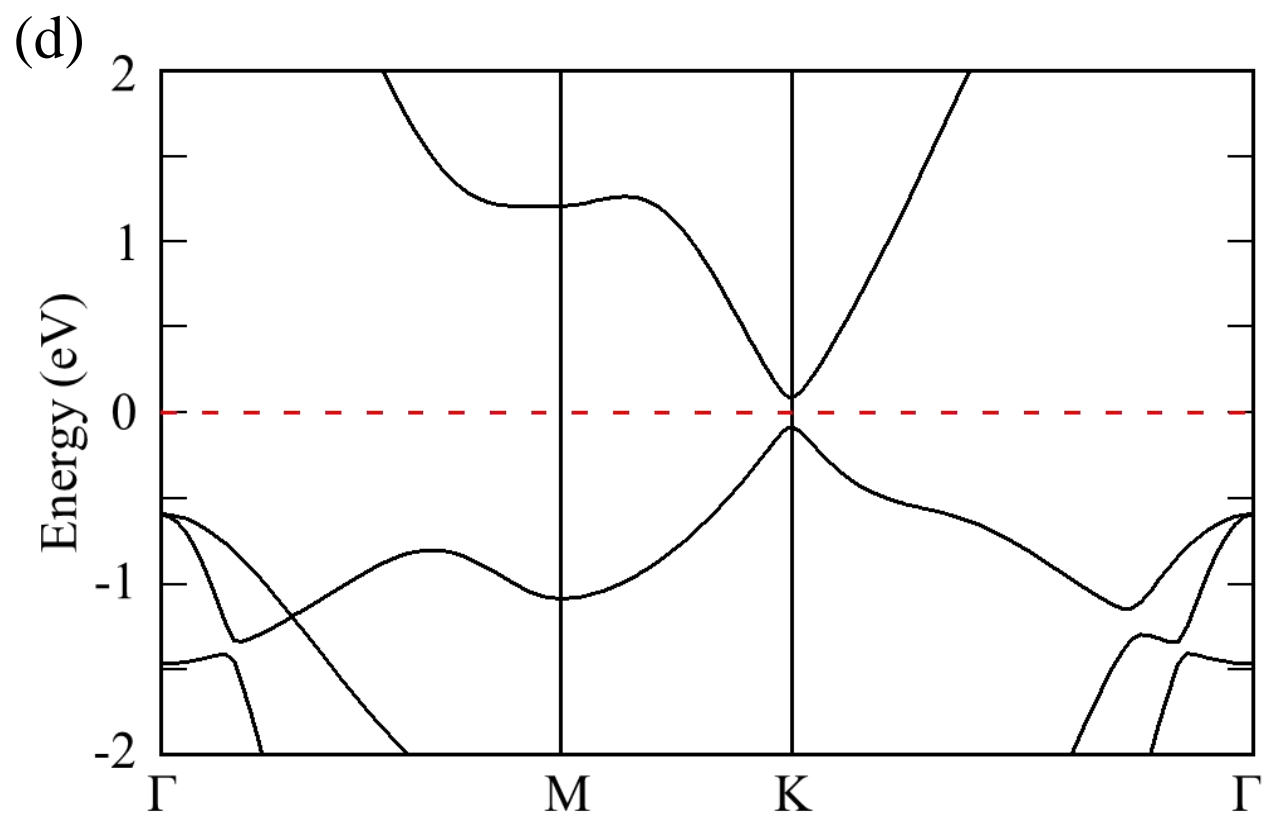

# Supplementary Materials

**Two-dimensional MX family of Dirac materials with tunable electronic and topological properties**

Yan-Fang Zhang[1,2,†], Jinbo Pan[2,†], Huta Banjade[2], Jie Yu[2], Tay-Rong Chang[3], Hsin Lin[4], Arun Bansil[5], Shixuan Du[1]*, Qimin Yan[2]‡

[1]Institute of Physics & University of Chinese Academy of Sciences, Chinese Academy of Sciences, Beijing, 100190, China

[2]Department of Physics, Temple University, Philadelphia, PA 19122, USA

[3]Department of Physics, National Cheng Kung University, Tainan 701, Taiwan

[4]Institute of Physics, Academia Sinica, Taipei, Taiwan

[5]Physics Department, Northeastern University, Boston, MA 02115, USA

***DFT calculation methods:*** Geometric relaxations and electronic property calculations of MX compounds were performed using the density functional theory with projector-augmented wave (PAW) potentials[1,2] as implemented in the VASP code[3]. A vacuum slab of 20 Å and a plane-wave basis set with an energy cutoff of 520 eV were used. A 10×10×1 Γ-centered k-point mesh was applied to sample the Brillouin zone. The Perdew−Burke−Ernzerh (PBE) functional was used to treat exchange−correlation effects within the generalized gradient approximation (GGA) in most cases[4]; selected computations using the screened hybrid functional of Heyd, Scuseria, and Ernzerhof (HSE06)[5,6] were also carried out. All structures were fully relaxed until the energy and force converged to $10^{-7}$ eV and 0.001 eV/Å, respectively. To explore the dynamical stability of our materials, phonon dispersion analysis was performed by using the density functional perturbation theory method as implemented in the Phonopy code[7] interfaced with VASP. In phonon calculations, which are based on the PBE, a convergence criterion of $10^{-8}$ eV for energy was used.

***Carrier mobility calculations:*** Carrier mobility was computed using a phonon-limited scattering model in which the carrier mobility of 2D materials can be expressed by the simplified relation[8,9]

$$\mu = \frac{e\hbar^3 \, C_{2D}}{k_B T m_e^* m_d (E_1^i)^2} \tag{1}$$

where, $m_e^* = \hbar^2/(\partial^2 E/\partial k^2)$ is the carrier effective mass in the propagation direction and $m_d = \sqrt{(m_{para}^* m_{perp}^*)}$ is the average effective mass, and $m_{para(perp)}^*$ is the effective mass parallel (perpendicular) to the propagation direction. $E_1^i$ is deformation potential constant defined by $E_1^i = \Delta V_i/(\Delta l/l_0)$. Here, $\Delta V_i$ is the energy change of valence-band minimum for holes or the conduction band maximum for electrons along the propagation direction. $l_0$ and $\Delta l$ denote that lattice constant in the propagation direction and the deformation of $l_0$ under strain. $C_{2D}$ is the 2D elastic modulus, which can be obtained from the expression: $\Delta E/S_0 = C_{2D}(\Delta l/l_0)^2/2$, where $\Delta E$ is the total energy change of the unit-cell under strain in the propagation direction and $S_0$ is the area of unit-cell under equilibrium. The temperature T was set at 300 K.

We used Be *s* and *p*, and Ce *s* orbitals to construct Wannier functions via WANNIER90[10]. $Z_2$ topological invariant was calculated by the Wilson loop method[11].

***The correlation between work function and cation s orbitals***: According to our COHP analysis, the state at the *K* point is non-bonding, which implies that the absolute energy position of the Dirac cone and the Fermi energy is determined by the energy position of the *sp³*-like hybridized atomic orbital of the metal atoms. Since this hybridized orbital possesses strong *s* and *p$_z$* contributions, its energy involves both the position of the s-state (relative to vacuum) and the extent to which *s* and *p$_z$* states are hybridized. The compounds in which the cation species have lower *s*-orbital centers typically exhibit higher work functions. The *s*-orbital contribution also affects the work function because the cation *s*-orbitals are located at lower energies relative to the cation p-orbitals. A larger *s*-orbital contribution is thus associated with the appearance of the Dirac cone at a lower energy relative to the vacuum.

***s-orbital-center and orbital contribution calculations***: The *s*-orbital center is calculated by $\int g_s(E)EdE / \int g_s(E)dE$ and $\int g_p(E)EdE / \int g_p(E)dE$, where $g_s(E)$ and $g_p(E)$ are density of states of the *s*- and *p*-states of M atoms, respectively. The *s*-orbital contribution is obtained over the energy range -0.5 ~ 0.5 eV relative to the Dirac cone by computing: $\int_{Ef-0.5}^{Ef+0.5} g_s(E)dE / \int_{Ef-0.5}^{Ef+0.5} g_{tot}(E)dE$ and $\int_{Ef-0.5}^{Ef+0.5} g_p(E)dE / \int_{Ef-0.5}^{Ef+0.5} g_{tot}(E)dE$.

Table S1. Lattice constants and Fermi velocities for electrons and holes in MX monolayers.

|  | Lattice constant (Å) | electron velocity ($10^5$ m/s) | hole velocity ($10^5$ m/s) |
|---|---|---|---|
| **BeCl** | 3.14 | 6.64 | 4.76 |
| **BeBr** | 3.36 | 6.38 | 3.75 |
| **BeI** | 3.68 | 6.02 | 2.47 |
| **MgCl** | 3.60 | 6.72 | 6.89 |
| **MgBr** | 3.77 | 6.57 | 5.97 |
| **MgI** | 4.06 | 6.17 | 4.50 |
| **ZnCl** | 3.48 | 6.88 | 5.07 |
| **ZnBr** | 3.67 | 6.44 | 3.93 |
| **ZnI** | 3.94 | 5.92 | 2.50 |
| **CdCl** | 3.78 | 6.81 | 6.20 |
| **CdBr** | 3.94 | 6.41 | 4.65 |
| **CdI** | 4.18 | 5.90 | 3.25 |
| **graphene** | 2.46 | ~11.1 | ~10.4 |

Table S2. Predicted carrier mobilities of monolayer BeCl. *e* and *h* denote electron and hole carriers, respectively. $m_x^*$ and $m_y^*$ are carrier effective masses along directions x and y, respectively. $E_{1x}$ ($E_{1y}$) and $C_{x(2D)}$ ($C_{y(2D)}$) are the deformation potentials and 2D elastic modulus for the x (y) direction. $\mu_{x(2D)}$ and $\mu_{y(2D)}$ are carrier mobilities along the x- and y- directions, respectively.

|  | *e* | *h* |
|---|---|---|
| $m_x^*/m_0$ | 0.054 | -0.060 |
| $m_y^*/m_0$ | 0.066 | -0.065 |
| $E_{1x}$ (eV) | -1.18 | -0.78 |
| $E_{1y}$ (eV) | -1.27 | -0.67 |
| $C_{x(2D)}$ (Jm$^{-2}$) | 61.12 | 61.12 |
| $C_{y(2D)}$ (Jm$^{-2}$) | 71.34 | 71.34 |
| $\mu_{x(2D)}$ (10$^5$ cm$^2$V$^{-1}$s$^{-1}$) | 2.90 | 5.71 |
| $\mu_{y(2D)}$ (10$^5$ cm$^2$/(V$^{-1}$s$^{-1}$)) | 2.39 | 8.33 |

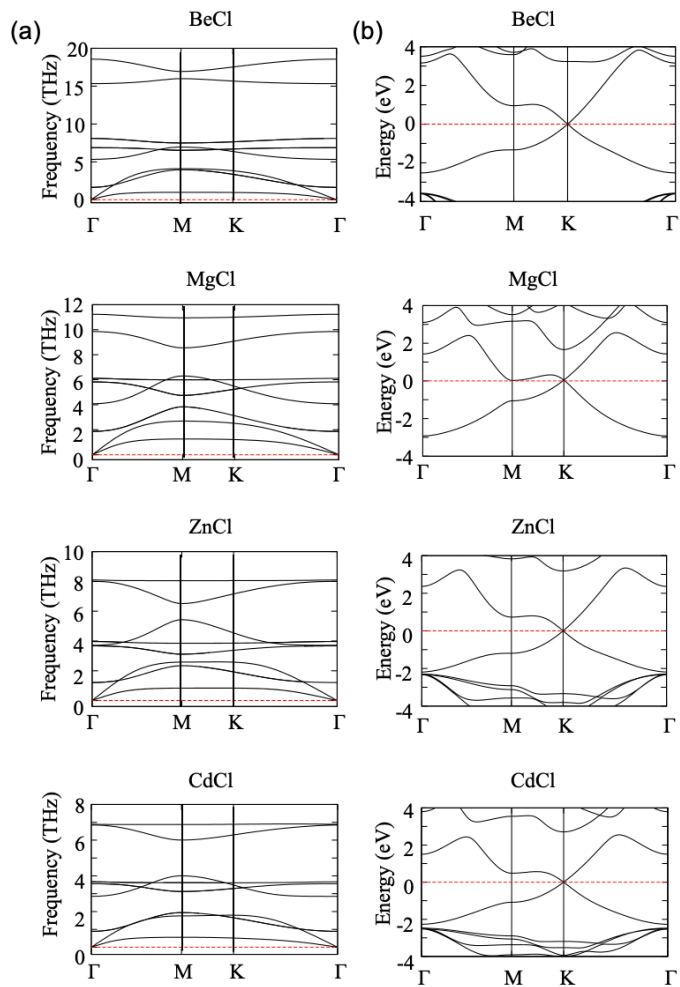

Figure S1. (a) From top to bottom: phonon-dispersions in BeCl, MgCl, ZnCl and CdCl. (b) From top to bottom: band structures of BeCl, MgCl, ZnCl and CdCl.

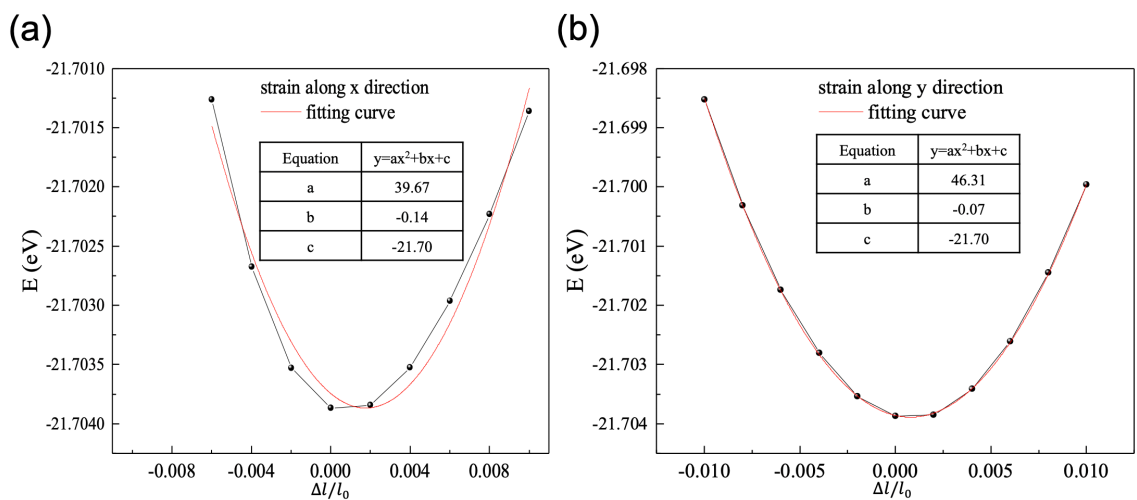

Figure S2. Relationship between total energy and strain along (a) x-direction and, (b) y-direction. The curve can be fitted as $y=ax^2+bx+c$, the $C_{2D}$ thus can be obtained from $2a/S_0$.

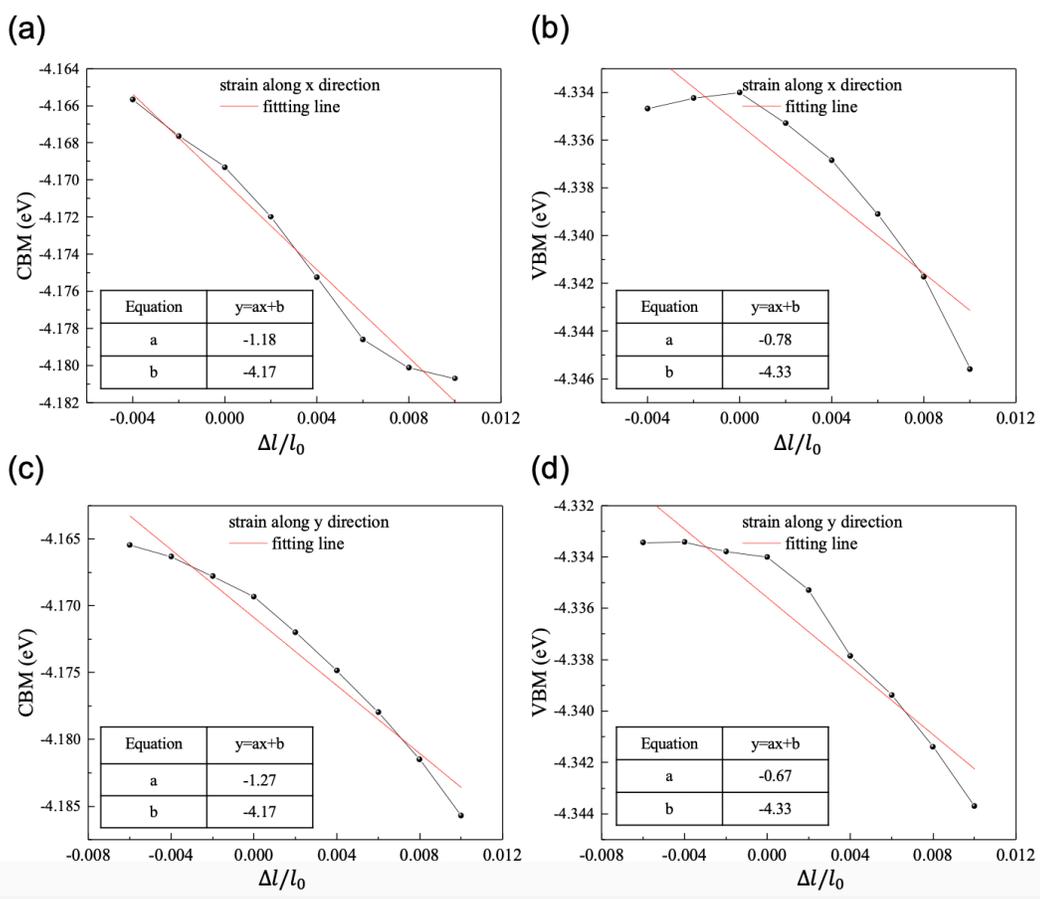

Figure S3. (a,b) Variations in the energies of CBM and VBM with strain along the x-direction. (c,d) Same as (a,b) except these panels refer to the y-direction. Lines are fitted as: y=ax+b, where a is the deformation potential.

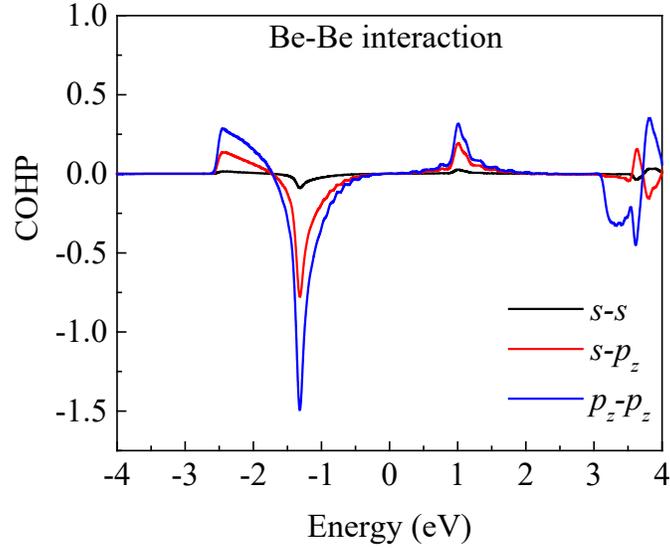

Figure S4. Crystal Orbital Hamiltonian Population (COHP) analysis of orbital interactions close to the Dirac cone (located at the energy zero in the figure). Be-$s$/Be-$p_z$ and Be-$p_z$/Be-$p_z$ bonding interactions (COHP < 0) are seen immediately below the Dirac cone at the $K$ point, while antibonding interactions (COHP > 0) are seen in the band above the Dirac cone.

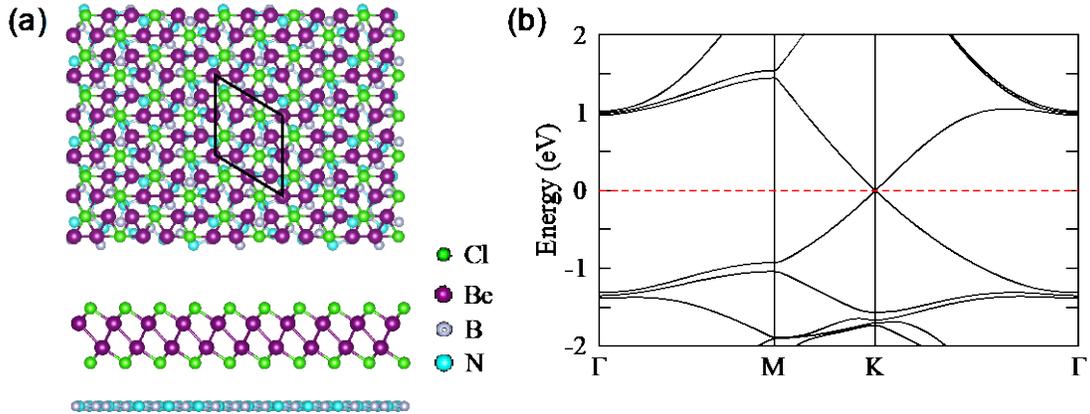

Figure S5. Geometric and electronic structures of BeCl on $h$-BN. (a) Top and side views of BeCl on $h$-BN. The green, purple, silver and cyan balls represent Cl, Be, B and N atoms, respectively. (b) The band structure of BeCl on $h$-BN.

*Effects of substrate:* Single crystal $h$-BN on a wafer scale has been recently successfully fabricated and is confirmed as a good substrate for the epitaxial growth of transition metal dichalcogenides[12], while a monolayer MX compound can be possibly synthesized using a similar method. To evaluate the effect of the substrate on the electronic structure, we constructed a 2 × 2 superlattice of BeCl stacked on a $\sqrt{7} \times \sqrt{7}$ superlattice of $h$-BN with a lattice mismatch of 6.05%. The optimized structure is shown in Fig. S5(a). Geometry relaxation indicates that BeCl can be stabilized on the *h*-BN substrate without bending. The vertical distance between the bottom Cl atoms and the *h*-BN plane is around 3.35 Å, suggesting a van der Waals interaction between the monolayer BeCl and the *h*-BN substrate. Fig. S5(b) shows that the Dirac-cone electronic structure of BeCl is preserved at the Fermi energy with no overlapping band from the *h*-BN substrate. These results suggest the possibility that the Dirac-cone related exotic phenomena in this set of novel quantum materials will be amenable to exploration through spectroscopic techniques such as the angle-resolved-photoemission and scanning-tunneling spectroscopies.

***Electronic and circular polarization in Be$_2$ClI:*** The electrostatic potential distribution presented in Fig. S6(a) shows electronic polarization in Be$_2$ClI. Due to the existence of an energy gap in Be$_2$ClI, optical excitations can occur at the K and K′ valleys. This is reminiscent of the valley-selective circular dichroism found in MoS$_2$ and VSSe[13,14]. In order to explore this possibility, we calculated the degree of circular polarization in Be2ClI. The computed circular polarization (Fig. S6(b)) shows significant circular dichroism polarization at the K and K′ valleys.

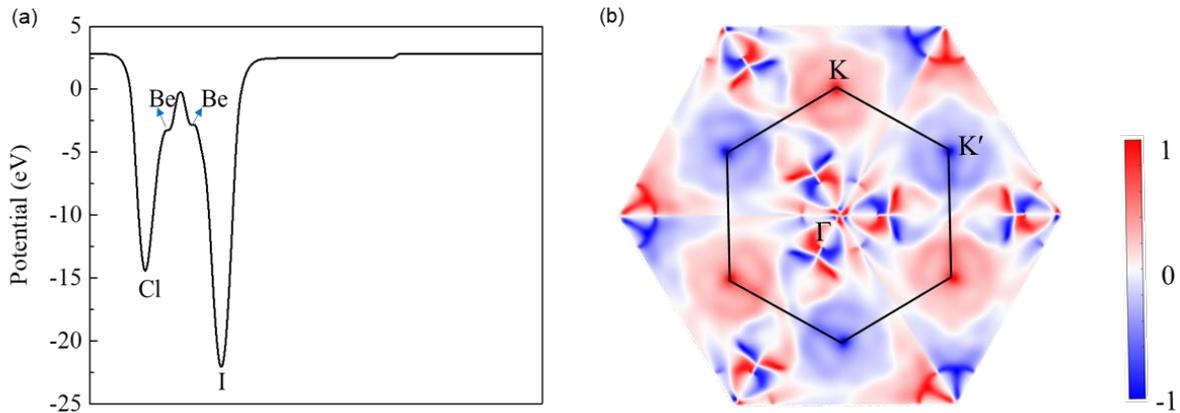

Figure S6 (a) Electrostatic potential distribution along the *z* direction. (b) Valley driven circular polarization. The black hexagon marks the first Brillouin zone.

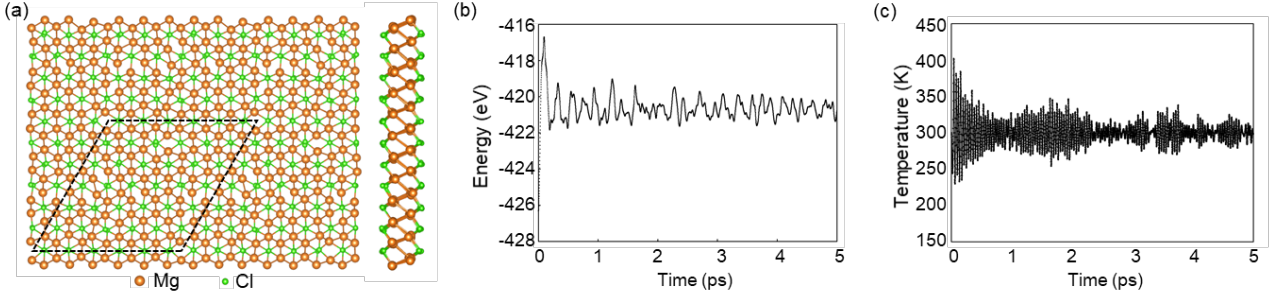

Figure S7. (a) Top and side views of the snapshot of MgCl crystal at the end of MD simulations at 300 K. The black rhombus marks the supercell used in the computations. The orange and green balls represent Mg and Cl atoms, respectively. (b) and (c) Evolution of temperature and system energy with time obtained from the MD simulation of MgCl crystal at 300 K.

***Tight-binding analysis:*** By considering only the nearest-neighbor hopping terms, the tight-binding (TB) Hamiltonian for *s and $p_z$* electrons in a simplified MX system (a graphene-like honeycomb lattice, see Fig. S8) can be written as follows.

$$H = \sum_{\langle i,j \rangle, \mu, \nu} t_{i,j,\mu,\nu} \left( a_{i,\mu}^\dagger b_{j,\nu} + H.c. \right),$$

where $a_{i,\mu}$ ($a_{i,\mu}^\dagger$) annihilates (creates) an electron with orbital $\mu$ ($\mu = s, p_z$) at site $R_i$ in the sublattice A (an equivalent definition is used for sublattice B), and $t_{i,j,\mu,\nu}$ is the nearest-neighbor hopping integral. We choose $\{s^A, p_z^A, s^B, p_z^B\}$ as the orbital basis to construct the Hamiltonian matrix. The resulting Hamiltonian in the k-representation is:

$$H_k = \begin{pmatrix} \varepsilon_s & h_{sz}^{AA} & h_{ss}^{AB} & h_{sz}^{AB} \\ h_{zs}^{AA} & \varepsilon_z & h_{zs}^{AB} & h_{zz}^{AB} \\ h_{ss}^{AB*} & h_{sz}^{AB*} & \varepsilon_s & h_{sz}^{BB} \\ h_{zs}^{AB*} & h_{zz}^{AB*} & h_{zs}^{BB} & \varepsilon_z \end{pmatrix},$$

where *s* and *z* denote $s$ and $p_z$ orbitals and A and B denote sublattice sites, respectively. The matrix elements are:

$$h_{ss}^{AB} = t_1 \left( e^{i\vec{k} \cdot \vec{\delta}_1} + e^{i\vec{k} \cdot \vec{\delta}_2} + e^{i\vec{k} \cdot \vec{\delta}_3} \right),$$
$$h_{sz}^{AB} = t_3 \left( e^{i\vec{k} \cdot \vec{\delta}_1} + e^{i\vec{k} \cdot \vec{\delta}_2} + e^{i\vec{k} \cdot \vec{\delta}_3} \right),$$
$$h_{zs}^{AB} = h_{sz}^{AB*}$$
$$h_{zz}^{AB} = t_2 \left( e^{i\vec{k} \cdot \vec{\delta}_1} + e^{i\vec{k} \cdot \vec{\delta}_2} + e^{i\vec{k} \cdot \vec{\delta}_3} \right),$$
$$h_{sz}^{AA} = h_{sz}^{BB} = t_4$$

where $\varepsilon_s, \varepsilon_z, t_1, t_2, t_3$ and $t_4$ are the on-site and nearest-neighbor electron hopping parameters between the orbitals contributing to the Dirac cone of MX (Table S3). $\delta_1$, $\delta_2$, and $\delta_3$ are the three nearest-neighbor vectors in real space given by:

$$\delta_1 = \frac{a}{2}(-\sqrt{3}, 1, 2\tan\theta), \delta_2 = \frac{a}{2}(\sqrt{3}, 1, 2\tan\theta), \delta_3 = -a(0, 1, \tan\theta).$$

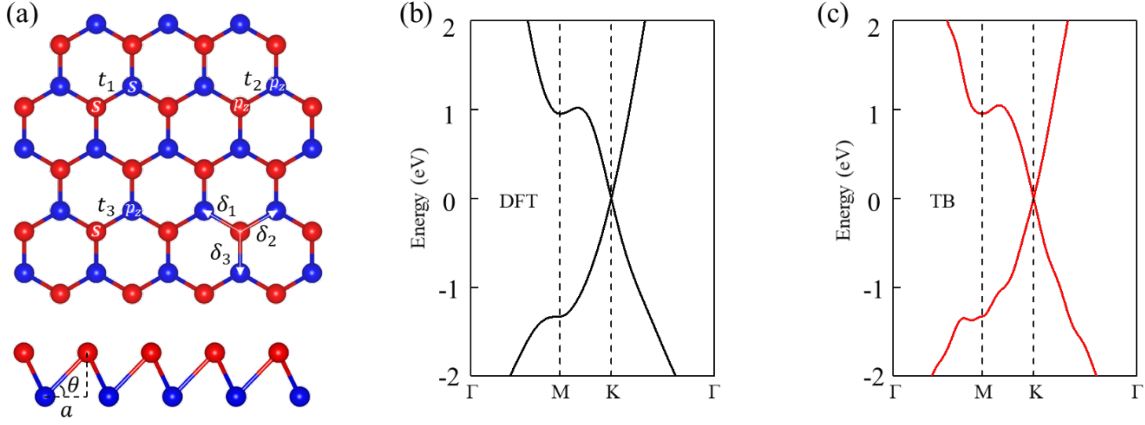

Figure S8. (a) Top and side views of the simplified MX system in which the hopping integrals between the atomic orbitals are indicated with $t_1, t_2$ and $t_3$. The red and blue balls describe the A and B sublattices, respectively. The nearest-neighbor distances between sublattices $\delta_1$, $\delta_2$, $\delta_3$ and the definition of buckling angle θ are also labeled. (b), (c) BeCl band structures obtained from DFT calculation and TB method, respectively.

Table S3: On-site and nearest-neighbor electron hopping parameters between the orbitals contributing to the Dirac cone of BeCl (in eV).

|       | s      | $p_z$  | s      | $p_z$  |
|-------|--------|--------|--------|--------|
| s     | -27.0  | 0.596  | 0.032  | 0.086  |
| $p_z$ | 0.596  | -1.162 | -0.086 | 0.310  |
| s     | 0.032  | -0.086 | -27.0  | -0.596 |
| $p_z$ | 0.086  | 0.310  | -0.596 | -1.162 |